\documentclass[twocolumn, aps, superscriptaddress, pra, citesort]{revtex4}

\usepackage{amssymb,amsmath}
\usepackage[pdftex]{graphicx}
\newcommand{\bra}[1]{\langle\left.{#1}\right|}
\newcommand{\ket}[1]{\left|{#1}\right.\rangle}
\newcommand{\braket}[1]{\langle{#1}\rangle}
\usepackage{comment}
\usepackage{bbm}
\usepackage{color}
\usepackage[toc,page]{appendix}
\newcommand{\SKIP}[1]{{\color{green} }} % #1  (insert ``#1'' or leave out) %\em for recognition of skipped section inside emacs

\begin{document}
\title{Chebyshev expansion for Impurity Models using Matrix Product States}

\author{Martin Ganahl}
\affiliation{Institut f\"ur Theoretische Physik, Technische Universit\"at Graz, 8010 Graz, Austria}
\author{Patrik Thunstr\"om}
\affiliation{Institute of Solid State Physics, Vienna University of Technology, 1040 Vienna, Austria}
\author{Frank Verstraete}
\affiliation{Faculty of Physics, University of Vienna, Boltzmanngasse 5,  1090 Vienna, Austria}
\affiliation{Department of Physics and Astronomy, Ghent University, Ghent, Belgium}
\author{Karsten Held}
\affiliation{Institute of Solid State Physics, Vienna University of Technology, 1040 Vienna, Austria}
\author{Hans Gerd Evertz}
\affiliation{Institut f\"ur Theoretische Physik, Technische Universit\"at Graz, 8010 Graz, Austria}

\begin{abstract}
We improve a recently developed expansion technique for calculating real frequency spectral functions of any one-dimensional model 
with short-range interactions,
by postprocessing computed Chebyshev moments with linear prediction.
This can be achieved at virtually no cost and, in sharp contrast to existing methods based on the dampening of the 
moments, improves the spectral resolution rather than lowering it. 
We validate the method for the exactly solvable resonating level model and the single impurity Anderson model.
It is capable of resolving sharp Kondo resonances, 
as well as peaks within the Hubbard bands when employed as an impurity solver for dynamical mean-field theory (DMFT).
Our method works at zero temperature and allows for arbitrary discretization of the bath spectrum.
It achieves similar precision as the dynamical density matrix renormalization group (DDMRG), at lower cost.
We also propose an alternative expansion, of $\mathbbm{1}-\exp(-\tau H)$ instead
of the usual $H$, which opens the possibility of using established methods for the time evolution of matrix product 
states to calculate spectral functions directly. 

\end{abstract}
\date{\today}
\maketitle

%=============================================
\section{Introduction}
%=============================================
For one-dimensional (1-d) strongly correlated quantum systems, the density matrix renormalization group (DMRG) 
\cite{white_density-matrix_1992,white_density-matrix_1993}
and matrix product states (MPS) in general \cite{schollwoeck_density-matrix_2010, verstraete_mps_2008}
have emerged as a powerful tool for the calculation of ground state and excited state properties.
Since its invention, the DMRG has been extended to treat dynamical correlation functions \cite{kuhner_dynamical_1999, jeckelmann_dynamical_2002} 
as well as real time evolution \cite{vidal_efficient_2004,vidal_efficient_classical_2004,daley_time-dependent_2004, white_real-time_2004},
and nowadays is considered the method of choice for tackling 1-d quantum systems.

Regarding  spectral functions, 
the first  attempt with DMRG involved a continued fraction expansion \cite{hallberg_density-matrix_1995}. This method failed  however
to produce reliable results for large systems. A major improvement was the introduction of the correction vector (CV) method 
\cite{kuhner_dynamical_1999} and its variational dynamical DMRG (DDMRG) formulation \cite{jeckelmann_dynamical_2002}. Both methods
are known to give highly accurate spectral functions for 1-d systems. They suffer however from two major drawbacks: first, one has to invert a large, non-hermitian and possibly ill-conditioned 
system of equations in a DMRG-like fashion, and second, one has to do full DMRG-like calculations for every single $\omega$ value. 
A similar approach has also been 
proposed in \cite{peters_spectral_2011}, where the CV method was used as an impurity solver within dynamical mean field theory 
(DMFT) \cite{Metzner89a,Georges92a,georges_dynamical_1996,Georges2004} for a multi-orbital system.

More recently, the continued fraction expansion has been combined with an MPS parametrization of the Krylov vectors \cite{dargel_adaptive_2011, dargel_lanczos_2012},
yielding decreased computational cost as compared to the DDMRG \cite{jeckelmann_dynamical_2002} method 
while giving results of comparable accuracy. 
In other recent work \cite{holzner_chebyshev_2011}, MPS methods
were combined with the Chebyshev expansion technique of the Kernel polynomial method (KPM) \cite{weisse_kernel_2006} to obtain highly accurate spectral functions for the isotropic Heisenberg model.

In the present paper, we propose to extend the KPM by postprocessing computed Chebyshev moments with linear prediction, which drastically improves the spectral resolution
while the Chebyshev moments are accessible with far lower computational effort compared to DDMRG \cite{holzner_chebyshev_2011}.
We also extend the approach  to treat interacting quantum impurity problems and implement a DMFT self-consistency cycle with the improved KPM as an impurity solver.
Our approach avoids the simultaneous targeting of ground state and excited-state necessary in DDMRG, which may be especially important for multi-orbital impurity solvers.
For the single impurity Anderson model (SIAM) \cite{anderson_localized_1961}, we obtain accurate results over a broad parameter range. 
For the DMFT, our results show a sharp peak within the Hubbard bands
of the Hubbard model in the vicinity of the Mott-Hubbard transition.
We also introduce an expansion in powers of $\mathbbm{1}-\exp(-\tau H)$ instead of $H$, which completely avoids an energy truncation
necessary in the original KPM method and leads to improved stability.
In general, the advantages of the proposed Chebyshev expansion of MPS are:
(i) the spectral function can be calculated directly for real frequencies, also at zero temperature; 
(ii) the flexibility to arbitrarily discretize the hybridization function allows for  good energy resolution at all frequencies,
     and results as precise as DDMRG;
(iii) the method is applicable not only to impurity models, but to any 1-d model with short-range interactions.

%=============================================
\section{Methods and Models}
%=============================================
\subsection{Kernel Polynomial Method}\label{sec:kpm}
%=============================================
The kernel polynomial method \cite{silver_kpm_1994,silver_kpm_maxent_1997,weisse_kernel_2006} is a numerical method
for expanding Greens functions $G(k,\omega)$ and spectral functions $A(k,\omega)$ of many-body quantum systems in orthogonal 
Chebyshev polynomials $T_n(\omega)=\cos(n\textrm{acos}(\omega))$.
To make this document self contained, we will in the following describe the basic properties of 
the KPM. In the mathematical literature, two types of Chebyshev polynomials are used: those of first and those of second kind. For the spectral function, we will only need those of the first kind which will  be called Chebyshev polynomials henceforth. 

For a quantum system with Hamiltonian $H$ at temperature $T=0$, the spectral function for the unoccupied part ($A^+(\omega)$) and occupied part ($A^-(\omega)$) of the spectrum has the form
\begin{align}\label{eq:spectral_function}
  A^+(\omega)=\bra{\Psi_0}c\,\delta\big(\omega -(H-E_0)\big)\,c^{\dagger}\ket{\Psi_0},\\
  A^-(\omega)=\bra{\Psi_0}c^\dagger\,\delta\big(\omega -(H-E_0)\big)\,c\ket{\Psi_0},
\end{align}
where we shifted the (non-degenerate) ground state $\ket{\Psi_0}$ to zero energy; $c$ and $c^{\dagger}$ are bosonic or fermionic annihilation and creation 
operators, respectively. The Chebyshev expansion converges only in the interval $[-1,1]$, since the Chebyshev polynomials $T_n(\omega)$ are unbounded as a function of their order $n$ for all $|\omega|>1$. 
The Hamiltonian therefore has to be rescaled by some factor $a$, such that the {single particle excitation energies} 
are moved into the interval $[-1,1]$, 
\begin{align}
  H\rightarrow\tilde H\equiv \frac{H-E_0}{a}.
\end{align}
Henceforth, we assume that $H$ has been rescaled to $\tilde H$
(see also sections \ref{sec:implementation} and \ref{sec:exp_tau}).

By inserting a representation of the Dirac delta function in terms of the orthogonal Chebyshev polynomials
\begin{align}\label{eq:dirac_delta}
  \delta(\omega-\tilde H)=\frac{1}{\pi\sqrt{1-\omega^2}}\left(1+2\sum_{n=1}^{\infty}T_n(\tilde  H) T_n(\omega)\right)
\end{align}
into Eq.~(\ref{eq:spectral_function}), one arrives at 
\begin{eqnarray}
  A^+(\omega)&=&\frac{1}{\pi\sqrt{1-\omega^2}}\left(\underbrace{\braket{\Psi_0|cc^{\dagger}|\Psi_0}}_{\mu^+_0}\right.\nonumber\\ &&\left.+2\sum_{n=1}^{\infty}\underbrace{\bra{\Psi_0}c\,T_n(\tilde H)c^{\dagger}\ket{\Psi_0} }_{{\mu^+_n}}T_n(\omega)\right).\label{eq:aplus}
\end{eqnarray}
The method amounts to computing the expectation values $\mu^+_n=\bra{\Psi_0}c\, T_n(\tilde H)\,c^{\dagger}\ket{\Psi_0}$ of the $n$-th Chebyshev polynomial. For 
many-body systems, this is of course a highly non-trivial task.

The Chebyshev polynomials satisfy the recursion relation
\begin{align}
  T_0(\tilde H)&=\mathbbm{1}\\
  T_1(\tilde H)&=\tilde H\nonumber\\
  T_{n}(\tilde H)&=2\tilde HT_{n-1}(\tilde H)-T_{n-2}(\tilde H).\nonumber
\end{align}
The computation of $\mu^+_n=\bra{\Psi_0}cT_n(\tilde H)c^{\dagger}\ket{\Psi_0}$ can therefore be performed through a corresponding
recursion relation for the many-body quantum states
\begin{align}
  \ket{t_0}&=c^{\dagger}\ket{\Psi_0}\label{eq:recurr}\\
  \ket{t_1}&=\tilde H\ket{t_0}\nonumber\\
  \ket{t_n}&=2\tilde H\ket{t_{n-1}}-\ket{t_{n-2}}\nonumber\\
  \mu^+_m&=\braket{t_0|t_m}\nonumber.
\end{align}
If $\tilde H$ has been properly rescaled, then this recursion relation will converge.
The product relations of the Chebyshev polynomials allow the moments $\mu_{2n}^+$ and $\mu_{2n+1}^+$ to be calculated already from the states $\ket{t_n}$ and $\ket{t_{n+1}}$ using \cite{weisse_kernel_2006}
\begin{align}\label{eq:recurr_2n}
\mu^+_{2n}=2\braket{t_n|t_n}-\mu^+_0\\\nonumber
\mu^+_{2n+1}=2\braket{t_{n+1}|t_n}-\mu^+_1.
\end{align}
If not stated otherwise, results in this paper were obtained from these reconstructed moments. 
The above procedure requires only the ability of applying an operator $\tilde H$ to a state $\ket{t_m}$ and of computing overlaps of the resulting states
with $\ket{t_0}$ or $\ket{t_{m-1}}$. The moments for the occupied part of the spectrum ($\mu^-$) can be generated by changing  $c^\dagger$ to $c$ in the first line of Eq. (\ref{eq:recurr}).
The full spectral function $A(\omega)$ is obtained by combining $\mu^+_n$ and $\mu_n^-$, and using $T_n(-\omega)=(-1)^n T_n(\omega)$:
\begin{align}
&A(\omega)= A^+(\omega)+A^-(-\omega)\label{eq:momentrelation}\\
&=\frac{1}{\pi\sqrt{1-\omega^2}}\left([\mu_0^++\mu_0^-]+2\sum_n[\mu_n^++(-1)^n\mu_n^-]T_n(\omega)\right)\nonumber\\
&=\frac{1}{\pi\sqrt{1-\omega^2}}\left(\mu_0+ 2\sum_n\mu_n T_n(\omega)\right)\nonumber
\end{align}
where $\mu_n \equiv \mu^+_n + (-1)^n \mu^-_n$. Note that the decay of the positive (negative) moments $\mu^{+}$ ($\mu^{-}$) with $n$ is qualitatively different from that of $\mu_n$:
The spectral function $A^+(\omega)$ ($A^-(\omega)$) has a step at the Fermi-energy $\omega=0$, which corresponds to an
 {\it algebraic} decay of $\mu^{+}$ ($\mu^{-}$) \cite{boyd_spectral_methods} of order 1 ($\mu^{+}\propto\frac{1}{n}$). 
The added moments $\mu_n$ on the other hand corresponds  to a smooth analytic spectral function for which the moments converge much faster (exponentially) to zero \cite{boyd_spectral_methods}.

%=============================================
\subsection{MPS implementation and energy truncation \label{sec:implementation}}
%=============================================
The recursion relation given by Eq. (\ref{eq:recurr}) can be implemented straightforwardly in 
an MPS framework \cite{holzner_chebyshev_2011}. For this purpose, the Hamiltonian 
$\tilde H$ is brought into a matrix product operator (MPO) form \cite{schollwoeck_density-matrix_2010,pirvu_mpo_2010}, 
formally similar to the MPS representation of a quantum state. The auxiliary dimension $D^{MPO}$ 
of the corresponding MPO-matrices is typically between 4 and 6. 
In general, the application of an MPO of bond dimension $D^{MPO}$ to an MPS of bond dimension $\chi$ (denoted $\ket{\chi}$ in the following)
leads to an MPS with increased bond dimension $\chi'=D^{MPO}\chi$. 
To make successive applications like in Eq.~(\ref{eq:recurr}) feasible, the state is then {\it compressed} by a variational procedure \cite{schollwoeck_density-matrix_2010} 
back to bond dimension $\chi$. This is the same procedure as done in standard DMRG calculations.
The corresponding systematic error is quantified by the fidelity $\epsilon=||\ket{\chi}-\ket{\chi'}||/||\ket{\chi'}||$
which measures the relative distance of the compressed and original state,
and can be estimated by the truncated weight, 
which is the sum of the discarded eigenvalues of the density matrix
\cite{schollwoeck_density-matrix_2010, verstraete_mps_2008}.

When $\tilde H$ has not been rescaled over the { full bandwidth} of $H$, compression 
reintroduces modes with energies outside the convergence interval ($|E| > 1$), which would result in a rapidly diverging recurrence. To overcome this divergence, an energy-truncation scheme has to be 
used to project out such high energy modes, at the cost of introducing a new systematic error and extra computational effort. 
In section \ref{sec:exp_tau} we present a generalization of the KPM method without any need for energy truncation.

Energy truncation is done similar to a DMRG run \cite{holzner_chebyshev_2011} by sweeping 
back and forth ($E_{sweep}$ times each) through the system. 
At each site, high energy modes are projected out by applying a projection operator. 
It is obtained by a Lanczos tri-diagonalization, which yields a set of approximate eigenenergies $E_n$ and 
eigenstates $\ket{E_n}$. 
The projection operator $\mathcal{P}$ projecting out modes with $|E_n|>1$ is then given by 
$\mathcal{P}=\mathbbm{1}-\sum_{|E_n|>1}^{D_{max}}\ket{E_n}\bra{E_n}$. $D_{max}$ is the number of steps in the Lanczos procedure.
For a detailed study on the effect of the $D_{max}$ on the accuracy of the moments $\mu_m$ see Ref.~\onlinecite{holzner_chebyshev_2011}. 
The appropriate size of $D_{max}$ depends on the rescaling parameter $a$ which determines the level-spacing of $H/a$. 
If the recurrence relation shows divergence, $D_{max}$ is increased until the recursion becomes stable.
Further runs with different $a$ and $D_{max}$ are required to ensure parameter-independence of the results.
We typically used $a$ equal to 10-20 times the bandwidth $2D$, 
and $D_{max}=5-30$.
When $a$ is chosen too  small, it cannot be compensated by increasing 
$D_{max}$ or $E_{sweeps}$, and the results become unstable.

An additional drawback of the energy truncation approach is that in contrast to ground state or compression algorithms, it is 
{\it not variational} in character, hence no notion of optimality can be associated with it, and convergence of the method is not guaranteed. % (see \ref{sec:exp_tau}). 
Energy truncation has been speculated to be the major limiting factor of accuracy \cite{holzner_chebyshev_2011}. 
In our calculations, we find that both usual matrix compression {\it and} energy truncation limit the accuracy of a simulation. 

%==============================================================
\subsection{Expansion of $1-\exp(-\tau H)$}\label{sec:exp_tau}
%==============================================================

For convergence of the Chebyshev recurrence, 
any one-to-one mapping $f(H)$ of the spectrum of $H$ into $[-1,1]$ is sufficient. 
A natural choice for $f(H)$ is to employ the exponential function, 
$\exp(-\tau (H-E_0+\epsilon))$, where $E_0$ is the ground state energy, and $\epsilon>0$
is a small energy shift which avoids getting too close to the boundary $f=1$.
Since the exponential function is bounded, no energy rescaling and no truncation step is needed.
Another advantage of this approach is that one can use 
a Trotter-decomposition of $\exp(-\tau (H-E_0+\epsilon))$, with sufficiently small $\tau$, which is a 
standard tool for solving time dependent many-body systems \cite{vidal_efficient_2004,vidal_efficient_classical_2004,daley_time-dependent_2004}.
For small $\tau$, $\exp(-\tau (H-E_0+\epsilon))\approx 1-\tau (H-E_0+\epsilon)$, and thus the spectral resolution is approximately constant. 

However, the positive and negative branches of the zero frequency peak of $A(\omega)$ now have to be 
calculated separately and then patched to give the full spectral function. 
A substantial drawback of this procedure is that both patches contain a jump at the Fermi energy $\omega=0$
(which is mapped to $\tilde \omega =1$). Using Eq.(\ref{eq:momentrelation}) on the moments $\tilde \mu_n^\pm$ of $f(H)$ results in 
a spectral function where the Fermi edge of the hole part is mapped to $\tilde \omega=-1$ and the Fermi edge of the
particle part is mapped to $\tilde \omega=1$. Thus, the resulting function has {\it two jumps}. The added moments $\tilde \mu_n$
then decay only algebraically, which requires many moments to be calculated and which is not well suited for linear prediction (see below).

The disadvantages are avoided by calculating the Chebyshev moments of $f(H)=\mathbbm{1}-\exp(-\tau (H-E_0))$.
Then the spectral function is smooth over the whole expansion interval. 
As a result, Eq.(\ref{eq:momentrelation}) can be used to good advantage and 
the full spectral function $\tilde A(\tilde\omega)$ of $f(H)$ 
can be obtained via moments $\mu_n=\mu_n^++(-1)^n\mu_n^-$.
It can be mapped back to $A(\omega)$ by plotting $\tau (1-\tilde \omega) \tilde A(\tilde \omega)$ 
vs.\ $-\ln(1-\tilde \omega)/\tau$.
We show initial results with this improved expansion in section \ref{sec:1-exp-tauH}.

%=============================================
\subsection{Linear Prediction}
%=============================================
Steps and sharp features of $A(\omega)$ will quite generally lead to ringing artefacts, known as Gibbs oscillations, due to the necessarily finite expansion order of the moments $\mu_n$. 
The usual remedy \cite{weisse_kernel_2006} is to multiply $\mu_n$ by damping factors $g_n$, i.e., $\hat \mu_n=\mu_n g_n$, and using $\hat \mu_n$ instead of  $\mu_n$ in Eq.~(\ref{eq:momentrelation}). Different damping factors $g_m$ are related to different constraints on the expansion of $A(\omega)$ (like causality, smoothness, and so on), and have been extensively discussed in the literature \cite{weisse_kernel_2006}. A common choice \cite{weisse_kernel_2006} is Lorentz damping 
\begin{equation}
g^L_n(\gamma)=\frac{\sinh\left(\gamma(1-\frac{n}{K})\right)}{\sinh(\gamma)},\label{eqn:lorentzdamp}
\end{equation}
where $K$ is the finite number of Chebyshev polynomials employed, and $\gamma$ is a parameter.

While removing unwanted  Gibbs oscillations to an extent depending on $\gamma$, this damping also leads to a reduction of spectral resolution. In the following we will present a different approach to correct Gibbs oscillations by numerically predicting the decay of the moments $\mu_m$, using a linear prediction algorithm \cite{white_spectral_2008,barthel_spectral_2009}.

Linear prediction is a simple yet powerful tool to predict the behavior of a time series of equidistant data points. 
It is based on the ansatz that a data point $x_n$ can be approximated by a fixed linear combination of the previous $L$ data points:
\begin{align}
x_n \approx \tilde{x}_n\equiv-\sum_{j=1}^L a_j x_{n-j}.\label{eq:prediction}
\end{align}
The fixed coefficients $\{a_j\}$ are obtained (``trained'') by minimizing the cost function
\begin{align}
  \mathcal{F}=\sum_{n=L+1}^{T} w_n  |\tilde x_n-x_n|^2 \label{eq:linpred},
\end{align}
using a {\em training window} of $T$ known data points. Here, $w_n$ is a weighting function which we choose to be constant. 
The minimizing condition $\nabla_{\bf a^*}\mathcal{F}=0$ yields a set of linear equations, also known as the normal equations:
\begin{align}
  R{\bf a}&=-{\bf r},\\ \nonumber
  R_{ij}&=\sum_{n=L+1}^{T}  w_n x^*_{n-i}x_{n-j},&\quad r_i=\sum_{n=L+1}^{T} w_n x^*_{n-i}x_n \,,
\end{align}
with $1\le i,j \le L$.
The coefficients in ${\bf a}$ are obtained by inverting the matrix $R$, i.e., in vector notation ${\bf a}=R^{-1}{\bf r}$. 
For reasons of numerical stability of the algorithm, we use a pseudo-inverse with 
a cutoff $\delta$ instead of the full inverse of $R$. Once the coefficients $a_j$ have been found, 
the data points at $L+k$ ($k>0$) can be predicted as 
\begin{equation}
\tilde x_{L+k}= \sum_{j=1}^{L} [M^k]_{1\,j} \, x_{L+1-j} \,, \label{eq:predictlp}
\end{equation}
where  
\[M=\left(\begin{array}{ccccc}
-a_1 & -a_2 & -a_3 & \dots & -a_L\\
1   & 0     &   0  & \dots & 0\\
0   & 1     &   0  & \dots & 0\\
\vdots   & \ddots  & \ddots &\ddots & \vdots\\
0  & 0       & \dots &1 & 0\\
\end{array}\right).
\]
Eq. (\ref{eq:predictlp}) can be reexpressed using a diagonal matrix $\lambda$ containing the eigenvalues $\lambda_i$ of $M$,
\begin{align}
M = U \lambda U^{-1},\nonumber\\
b_i = \sum_{j=1}^{L} U^{-1}_{i~j} x_{L+1-j},\nonumber\\
\tilde x_{L+k} =  [U \lambda^k \mathbf{b}]_1.\label{eq:predictlambda}
\end{align}
From the last line in Eq. (\ref{eq:predictlambda}) it is clear that the sequence of predicted data points will diverge if any $|\lambda_i|>1$. These divergences can arise due to numerical inaccuracies in 
the training moments, or when the spectral function has some weight outside the interval $[-1,1]$. In such cases these eigenvalues can be either set to zero or,
as done in the present paper, 
renormalized to unity by \mbox{$\lambda_i\rightarrow\lambda_i/|\lambda_i|$}. 
The choice should not matter as long as the corresponding coefficient $b_i$ is small. 

Eq. (\ref{eq:predictlambda}) also shows that linear prediction is best suited to reproduce 
time series (which may contain oscillations) with an exponentially decaying envelope. 
It is therefore advantageous to use prediction on the added moments 
$\mu=\mu^++(-1)^n\mu^-$, which will indeed decay exponentially 
when $A(\omega)$ has no singularities in the expansion interval \cite{boyd_spectral_methods} (see above), 
rather than on $\mu^+$ and $\mu^-$ separately. Similar to Ref.\ \cite{barthel_spectral_2009}, 
we subdivided our data as $L=T/2$, which we found to give stable and accurate results.

%=============================================
\subsection{Single Impurity Anderson Model}
%=============================================
In general, an impurity model consists of a local interacting quantum system which is in contact with an infinite bath of non-interacting degrees of freedom, typically fermionic ones. In this paper we will focus on the single impurity Anderson model (SIAM) \cite{anderson_localized_1961}, an archetypal impurity model. 
It consists of a single impurity with interaction, immersed in a sea of non-interacting spin-half fermions, given by the Hamiltonian
\begin{equation}\label{eq:siam}
  H= \epsilon_f\!\sum_{\sigma}\!n_{0\sigma}+U n_{0\downarrow}n_{0\uparrow}+\sum_{k\sigma}\!\epsilon_kn_{k\sigma}+\!\sum_{k,\sigma}\!V_k c_{0\sigma}^{\dagger}c_{k\sigma}^{\phantom{\dagger}}+h.c. \,.
\end{equation}
Here, $U$ denotes the interaction, $\epsilon_k$ the energy-momentum dispersion of the bath,
$n_{k(0)\sigma} =c_{k(0) \sigma}^{\dagger}c_{k(0) \sigma}^{\phantom{\dagger}}$,
and $V_k$ is the hybridization between impurity states with creation operator $c_{0\sigma}^{\dagger}$
and bath states $k$ with annihilation operator $c_{k\sigma}$.
The impurity potential $\varepsilon_f$ contains the chemical potential $\mu$.

The effect of the bath can be fully described by the spectrum of the hybridization function $\Delta(\omega+i\eta)=\sum_k\frac{|V_k|^2}{\omega+i\eta-\epsilon_k}$, 
with an imaginary part $\tilde \Delta(\omega)\equiv -\frac{1}{\pi}\Im(\Delta(\omega))=\sum_k|V_k|^2\delta(\omega-\epsilon_k)$.

Eq.~(\ref{eq:siam}) can be mapped onto a chain geometry by discretizing this spectrum and, within each subinterval, expanding $c_{k\sigma}$ in plane waves \cite{bulla_nrg_review_2008}. 
For a logarithmic discretization mesh $E_n=\pm D\Lambda^{-n}$, where $D$ is the half bandwidth of the bath spectral function, this mapping can be done analytically (note that the discretization  
becomes exact only in the limit $\Lambda \rightarrow 1$) \cite{hewson}. In case of a $k$-independent hybridization $V_k=V$ and a flat, particle-hole symmetric bath-spectral function 
$\rho(\omega)=\sum_k\delta(\omega-\epsilon_k)=1/(2D)$ for $\omega\in[-D,D]$ 
($D=1$ unless stated otherwise), one obtains
\begin{align}\label{eq:siam_disc}
  H=&\epsilon_f\sum_{\sigma}n_{0\sigma}+U n_{0\downarrow}n_{0\uparrow} +\\
  +\sqrt{\xi_0}\sum_{\sigma}&(c_{0\sigma}^{\dagger}c_{1\sigma}+h.c.)+
  \sum_{\sigma, n=1}^{\infty}t_n(c_{n\sigma}^{\dagger}c_{n+1\sigma}+h.c.)\nonumber
\end{align}
where $\xi_0=V^2$ is the norm of $\tilde \Delta(\omega)$. 
$V$ determines the hybridization strength $\Gamma=\pi V^2\rho(0)$
and $t_n/D=(1+\Lambda^{-1})(1-\Lambda^{-n-1})\Lambda^{-n/2} / (2\sqrt{(1-\Lambda^{-2n-1})(1-\Lambda^{-2n-3})})$.

In order to make it amenable to a numerical treatment, the infinite chain is cut at finite length $N$, which is equivalent to a low-energy cutoff of the bath degrees of freedom.
For other hybridization functions and arbitrary discretizations, one has to resort to numerical techniques \cite{bulla_nrg_review_2008} with high precision arithmetics for the mapping of the 
higher dimensional impurity problem onto a chain geometry. 

The SIAM Hamiltonian in Eq.~(\ref{eq:siam_disc}) is the starting point for various numerical 
schemes \cite{nuss_variational_2012,gull_continuous_2011,haverkort_2014,weichselbaum_variational_2009,zitko_energy_resolution_2009,bulla_nrg_review_2008,raas_high-energy_2004} designed to compute ground state properties 
as well as dynamical properties of the impurity model,  the most famous one being Wilson's numerical renormalization group (NRG) \cite{Wilson75,bulla_nrg_review_2008}.
One of the most significant effects of a finite interaction is the redistribution of spectral weight of the impurity 
spectral function into three distinct features, the so-called upper and lower Hubbard satellites, and a zero-frequency peak, the Abrikosov-Suhl (or Kondo) resonance. The latter shows an exponentially 
decreasing width with increasing interaction, and determines the low-energy physics of the model. Though NRG yields highly accurate results for this low-energy part of the spectrum, the high-energy features 
of the spectral function are usually poorly resolved due to the small number of points used in the logarithmic bath discretization at high energies. 
Typical values of the discretization parameter range from  $\Lambda=1.5$ to $2$. Smaller values drastically increase the computational effort for finding the ground state of the system, 
since an increasing number of states has to be kept during the NRG-iterations, which eventually becomes impossible to continue.
Also, the central assumption of scale-separation of energies is no longer valid.
Using more sophisticated methods, the resolution at finite frequencies can be improved. 
In Refs.~\onlinecite{osolin_pade_2013,zitko_energy_resolution_2009}, the spectral function was obtained by averaging
over many different discretizations ($z$-averaging \cite{campo_2005}) in combination with using a very narrow broadening of the delta peaks obtained from NRG. This averaging procedure smoothens out the peaked structure of each single NRG-spectrum.

An advantage of MPS-based methods over NRG is the possibility to use an arbitrary discretization of the energy 
mesh for the bath spectral function, which can be used to increase the resolution of high energy features of the spectral function. In particular, the use of a linear instead of a logarithmic mesh at high energies helps to 
resolve high energy features of the spectral function. 

Another shortcoming of the NRG is the exponential increase of computational cost with the number of impurity orbitals. For an $N_p$-orbital model, every orbital couples to its own bath of free electrons. In NRG, all local degrees of freedom have to be treated as a single site, giving a scaling of $N(d^{N_p}\chi)^3$, 
where $d$ is the local Hilbert space dimension of a single orbital. In contrast, for MPS, 
a simple unfolding of the problem can reduce the complexity down to $N_pN(d\chi)^3$ for suitable models \cite{holzner_matrix_2008,saberi_matrix_2008}. 
For the SIAM, 
one can separate the two spin-degrees of freedom by unfolding the chain of spinful electrons into two chains of spinless fermions, interacting which each other at a single site.
All calculations in this paper have been obtained using such a mapping.

%=============================================
\subsection{Dynamical Mean Field Theory}
%=============================================

Dynamical mean field theory (DMFT) \cite{Metzner89a,Georges92a,georges_dynamical_1996,Georges2004} is a powerful method for the calculation of properties of strongly correlated models and materials. The central 
object of this theory is the local Greens function $G(\omega)$ of the full model at a given site. 
The basic idea of DMFT is to approximate the effect of the interacting lattice electrons surrounding a given site
by an appropriately chosen bath of free electrons at energies $\epsilon_{\nu}$ and hybridization of strength $V_{\nu}$ with the local site, yielding a hybridization function $\Delta(\omega_+\equiv \omega+i\eta)=\sum_{\nu}\frac{|V_{\nu}|^2}{\omega_+-\epsilon_{\nu}}$.
The lattice problem is thus mapped onto an {\it impurity} problem of SIAM type. 

In the DMFT self consistency cycle,
from the self-energy $\Sigma(\omega_+)$, the lattice Greens function $G(\omega_+)$ is calculated through the standard Dyson equation of the lattice.
Then a non-interacting impurity Greens function  $[\mathcal{G}_0(\omega_+)]^{-1}= \Sigma(\omega_+) + [G(\omega_+)]^{-1}$ defines a SIAM with hybridization  $\Delta(\omega_+)= \omega_+-\epsilon_f-[\mathcal{G}_0(\omega_+)]^{-1}$. We obtain the Greens function $G(\omega_+)$ of this SIAM by our MPS solver; from $G(\omega_+)$ a new  self-energy $\Sigma(\omega_+)= [\mathcal{G}_0(\omega_+)]^{-1} - [G(\omega_+)]^{-1}$ is calculated. With this new self-energy the self-consistency cycle is iterated until converegency.

As for the MPS implementation, let us note that the impurity Greens function can be obtained from the Chebyshev moments \cite{weisse_kernel_2006} through %  ( eq. (140) in \cite{weisse_kernel_2006})}
\begin{eqnarray}
  G^{imp}(\omega_+\equiv \omega+i \eta)&&=\frac{-i}{\sqrt{1-(\omega_+)^2}}\big(\mu_0+  \\ &&\!\!\!\!\!\!\!\!  2\sum_{n=1}^{\infty} \mu_n\exp(-in\arccos(\omega_+))\big).\label{eq:green}
\end{eqnarray}
In the calculations, the small imaginary shift $\eta$ acts as a regularization parameter. It is set to a small non-zero value ($\approx 10^{-5}$) 
to make sure that the spectral density remains positive even in the presence of small Gibbs-like oscillations at the band edges. 

In the present paper, we consider the Hubbard model on a Bethe lattice with infinite connectivity,
for which the DMFT gives the exact solution \cite{Metzner89a,georges_dynamical_1996}. 
In this special case, the new SIAM hybridization function can also be calculated  directly from the
last iteration's Green function:
\begin{align}
  \Delta(\omega_+)=\frac{D^2}{4}G(\omega_+),\label{eq:dmft_self_con_n}
\end{align}
where $D$ is half the bandwidth of the free lattice model.

%=============================================
\section{Results}
%=============================================

% {\bf Full list of parameters is:} \\
%     Physics:   U, $\Gamma$, D=1, $\epsilon_F=U/2$, shape of spectral function.\\
%     Mapping to chain: $\Lambda$, $N$.\\
%     MPS: $\chi$; for $\hat H$:  $a$, $D_{max}$, $E_{sweep}$.\\
%     Linear prediction: $\delta$, size(s) of training window, number of predicted moments ,($\gamma_{lorentz}$).}

% (All results for SIAM, $\epsilon_f=U/2$ (particle symmetric) and $D=1$ throughout)
%  D=1 except for DMFT section, half filling.

%=============================================
\subsection{Benchmark: Resonating Level Model\label{sec:rlm}}
%=============================================
As a first test for our method, we study the SIAM in the non-interacting limit ($U=0$), also known as the resonating level model (RLM),
which is exactly solvable. Each component of our method can therefore be benchmarked separately and the calculated quantities can be compared to exact results. For $U=0$ 
the Hamiltonian in Eq. (\ref{eq:siam}) contains only quadratic terms, which makes it diagonal in its single-particle eigenbasis. It is therefore easy to perform the recursion relation in Eq. (\ref{eq:recurr}) 
for finite systems of moderate size ($N \sim\mathcal{O}(10^2)$) to generate the exact Chebyshev moments to any order. 

Furthermore, for an infinite system, the local Greens function and its spectral function $A(\omega)$ can be 
computed analytically using an equation of motion approach \cite{bruus_flensberg}. For a flat density of states of the bath electrons and a constant 
hybridization $V$, the exact result is
\begin{align}
  A(\omega)=-\frac{1}{\pi}{\rm Im}\left( \frac{1}{\omega-\epsilon_f +\Delta(\omega)}\right)\label{eqn:rlm_aexact}\\
  \Delta(\omega)=\Gamma\left(i+\frac{1}{\pi}{\rm ln}\left(\frac{1-\omega/D}{1+\omega/D}\right)\right)\nonumber 
\end{align}
where $\Gamma=\pi V^2\rho(0)$ and $2D$ is the bandwidth of the bath spectral function\cite{zitko_energy_resolution_2009,bruus_flensberg}. The results for the RLM are obtained with 
such a bath, with  bandwidth $2D = 2$, $\Gamma = 0.005$, and $\epsilon_f=0$ (particle-hole symmetric point).
The moments of this function can be obtained using numerical integration, and will be referred to as $N=\infty$ results.

%=============================================
\subsubsection{Linear prediction}
\begin{figure} % Fig 1a, 1b
\includegraphics[width=1.02\columnwidth]{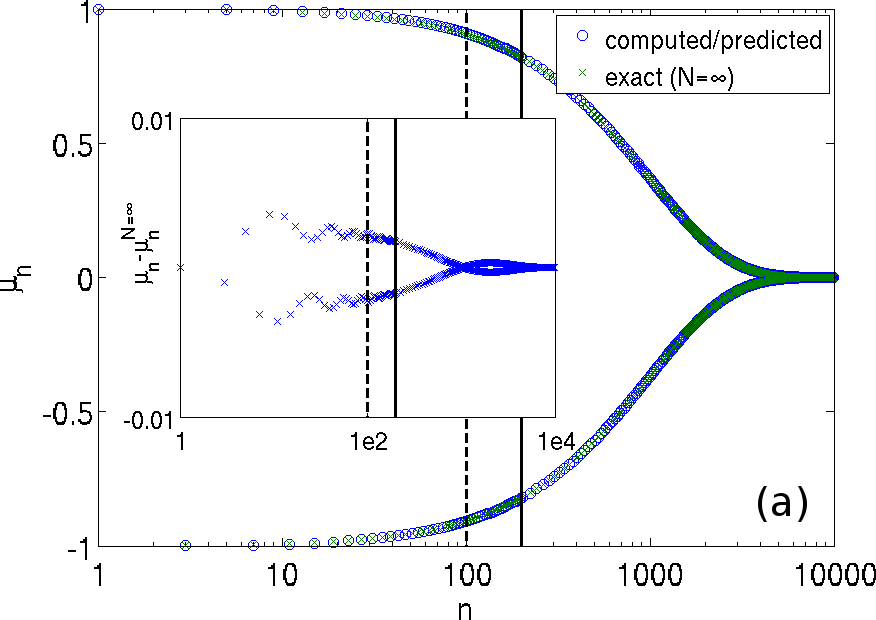}
  \includegraphics[width=0.98\columnwidth]{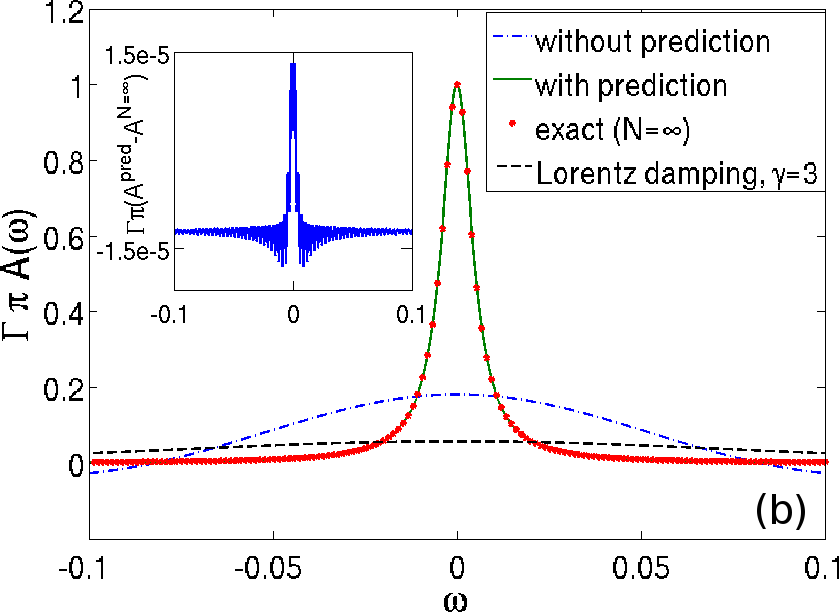}
  \caption{
   (Color online)
   (a) 
       Linear prediction using {\em exact} Chebyshev moments of an $N=100$ site RLM chain (circles), 
       compared to exact $N=\infty$ moments (crosses). Only even moments are plotted.
       The prediction was trained on the 100 moments between the two vertical bars.
       For better visibility, only every 20th moment is plotted at $n>200$ (note the logarithmic scale). 
       The inset shows the difference of the computed/predicted moments at $N=100$ 
       from the exact $N=\infty$ results.
       (Parameters: $\Lambda=1.05$, $\Gamma=0.005$, $\epsilon_f=0$, $\delta=10^{-5}$).
   (b) Spectral function without (dash-dotted blue line, using 200 moments) and with (solid green line) linear prediction. 
       The dashed black line shows results obtained with Lorentz damping. 
       The red dots represent the exact $N=\infty$ result (Eq. (\ref{eqn:rlm_aexact})), which 
       is very close to the results with linear prediction. The difference is shown in the inset.
   }\label{fig:rlm_exact}
\end{figure}

%=============================================
We start by comparing the moments obtained by linear prediction with exact moments. % ($N=\infty$). 
In Fig.~\ref{fig:rlm_exact}(a), 
200 moments to the left of the solid black line were calculated directly from Eq. (\ref{eq:recurr}), for an $N=100$ chain. The linear prediction algorithm was trained by predicting the 100 moments 
between the dashed and the solid line. 
Subsequently, we predicted 10000 moments. 
We note that the exact high order moments  for the finite $N=100$ chain would contain drastic finite size effects
(essentially from boundary reflections of the signal generated by applying $c^\dagger$).
We therefore compare the predicted moments to the exact $N=\infty$ ones.
Fig.~\ref{fig:rlm_exact}(a) shows that the predicted moments are very close to the exact ones, 
demonstrating the ability of the method to produce accurate 
results for Chebyshev moments. 
For the case $\epsilon_f\neq 0$ (not shown), where the decay of the moments is superimposed on oscillations, we get similar accuracy. 

Fig.~\ref{fig:rlm_exact}(b) shows the corresponding spectra. It should be noted that with increasing expansion order $K$, i.e, including more Chebyshev moments, the energy resolution of the KPM approximation 
improves like $1/K$. Linear prediction vastly increases the achievable resolution and also removes spurious oscillations that would result from a hard cutoff of the KPM approximation.
%-------------------------------
%-------------------------------

%=============================================
\subsubsection{MPS-computed moments}
\begin{figure} % Fig. 2
\includegraphics[width=1.02\columnwidth]{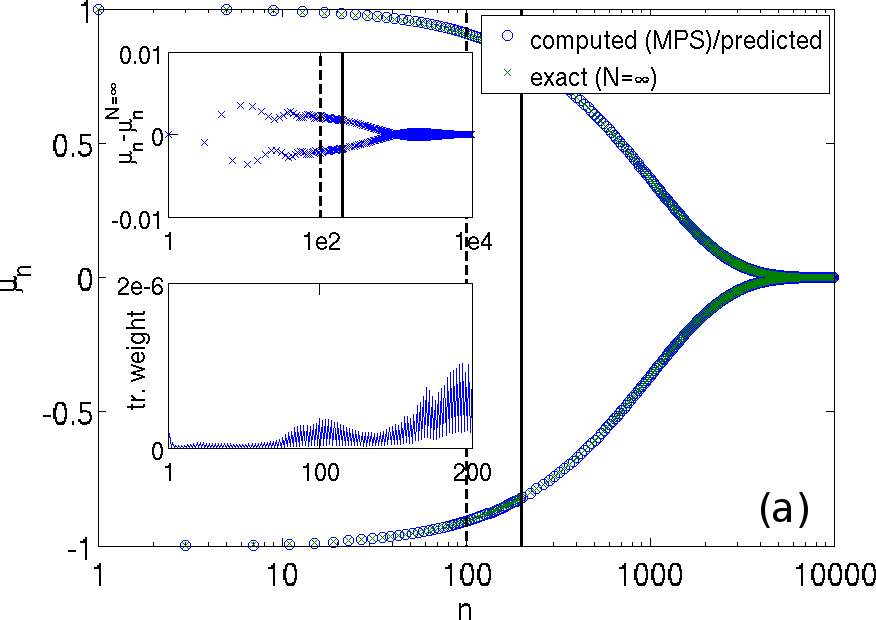}
    \includegraphics[width=0.98\columnwidth]{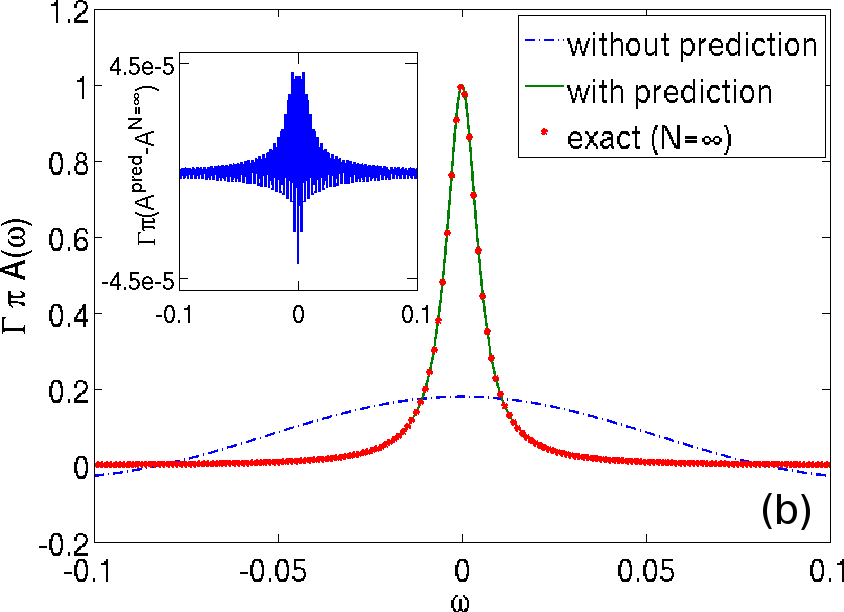}
\caption{
   (Color online)
  Same as Fig.~\ref{fig:rlm_exact}, but with {\em MPS computed} Chebyshev moments. 
       (MPS matrix dimension $\chi=250$, rescaling $a=5$, $D_{max}=20$, $E_{sweep}= 5)$ 
       Lower inset in (a): truncated weight of the first 200 MPS-computed moments.
\label{fig:rlm_mps}
  }
\end{figure}

%=============================================
We now turn to the MPS-computation of the Chebyshev moments \cite{holzner_chebyshev_2011}. 
The RLM is a non-trivial problem to MPS algorithms, 
even though it is exactly solvable, 
because of non-trivial entanglement between the orbitals of the chain (see Appendix). 
Finite entanglement gives rise to compression errors at finite matrix dimension,
and also to energy truncation errors (Sec. \ref{sec:implementation}). 
These errors can be estimated at each step in the iterative procedure, 
but to evaluate the overall error, including the effect of error cancellation, 
one needs the exact Chebyshev moments to compare with.
Fig.~\ref{fig:rlm_mps}(a) shows a comparison of the MPS-computed moments with exact ones ($N=\infty$).
The upper inset shows the difference between the exact and MPS-computed moments and the growth of the truncated weight, respectively. 
For the non-interacting 
RLM, the MPS method does in fact yield quasi-exact results. The lower inset shows the truncated weight for the first 200 moments.

We then predicted 10000 moments from the first 200 MPS-computed moments, and compare the resulting spectrum to the exact result given by Eq. (\ref{eqn:rlm_aexact}), 
as shown in Fig.~\ref{fig:rlm_mps}(b). 
Like in Fig.~\ref{fig:rlm_exact}(b), with the training moments alone it is not possible to properly resolve the sharp resonance at the Fermi energy. The results are on top of each other, demonstrating that 
linear prediction based on the MPS-calculation of 
200 moments essentially gives exact results for the RLM.
% (inset shows the difference to the Eq.(\ref{eqn:rlm_aexact})).

%--------------------------------------------
%--------------------------------------------

%=============================================
\subsection{Single Impurity Anderson Model\label{sec:siam}}
%=============================================
We now turn to the case of finite interaction strength $U > 0$, which renders the solution of Eq.~(\ref{eq:siam}) a highly non-trivial task. This situation is interesting both from a physical point of view and as a numerically demanding benchmark for our method. The calculations in this section are performed for a semicircular bath DOS with bandwidth $2D\equiv 2$, $\Gamma = 0.5$, and $\epsilon_f=-U/2$ (particle-hole symmetric point) in the regime $U \ge D$.
As a consequence of the large $U$, there is no conduction electron bath at the energy scale of the Hubbard bands, which results in extremely sharp Hubbard bands. 
A linear energy discretization corresponding to $N = 120$ sites 
\cite{bulla_nrg_review_2008} 
is used in the calculations throughout this section, to properly resolve all the spectral features. 
%The MPS calculations were performed with 
%$a = 12$, $\chi=200$, $D_{max}=25$, and $E_{sweeps}=5$. 
For prediction, we used a cutoff $\delta=10^{-6}$.

Additional benchmark calculations which focus on the more standard situation of a flat and wide ($D>U$) DOS and employ a logarithmic discretization can be found in the appendix. % \ref{app:l

%=============================================
\subsubsection{MPS-computed moments}
\begin{figure} % Fig. 3
  \begin{minipage}{1.01\columnwidth}
    \includegraphics[width=1\textwidth]{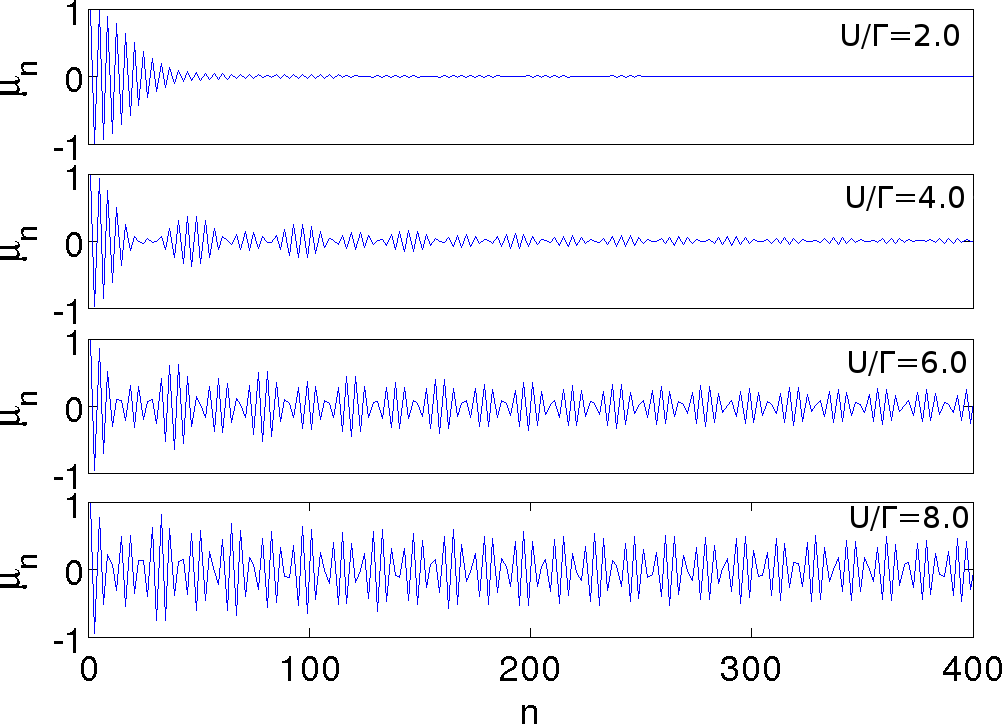}
  \end{minipage}
  \caption{(Color online) MPS-computed Chebyshev moments of the SIAM for $U/\Gamma=2,4,6,8$. 
    At large $U/\Gamma$, the moments show a much slower decay to zero. 
(Other parameters: $a = 12$, $\chi=200$, $D_{max}=25$, $E_{sweeps}=5$.)
}%
  \label{fig:moments_decay}
\end{figure}

%=============================================
In Fig.~\ref{fig:moments_decay} we plot the Chebyshev moments $\mu_n$ as obtained from MPS calculations for different values of the interaction strength $U/\Gamma=2,4,6,8$. For small $U/\Gamma\leq 4$, the 
moments decay to zero quickly, which indicates a rather featureless spectral function. In such cases, the moments obtained from the MPS calculations already produce  good resolution. For $U/\Gamma>4$ on 
the other hand, there is a slower decay, related to the emergence of sharp features in the spectral function \cite{raas_high-energy_2004} (see below); hence the linear prediction can significantly improve
the energy resolution for the impurity spectral function. For large values of $U/\Gamma$, the ground state of the system exhibits strong spin-fluctuations along the chain, resulting in a strong 
growth of the site-entanglement (see appendix). % \ref{app:ent}). 
In contrast to the  non-interacting limit and to the model studied in Ref.\ \onlinecite{holzner_chebyshev_2011}, for the SIAM this entanglement
can give rise to serious truncation errors.

%--------------------------------------------
%--------------------------------------------

%=============================================
\subsubsection{Linear prediction}
\begin{figure} % Fig. 4 a,b
  \begin{minipage}{1\columnwidth}
    \includegraphics[width=1.01\textwidth]{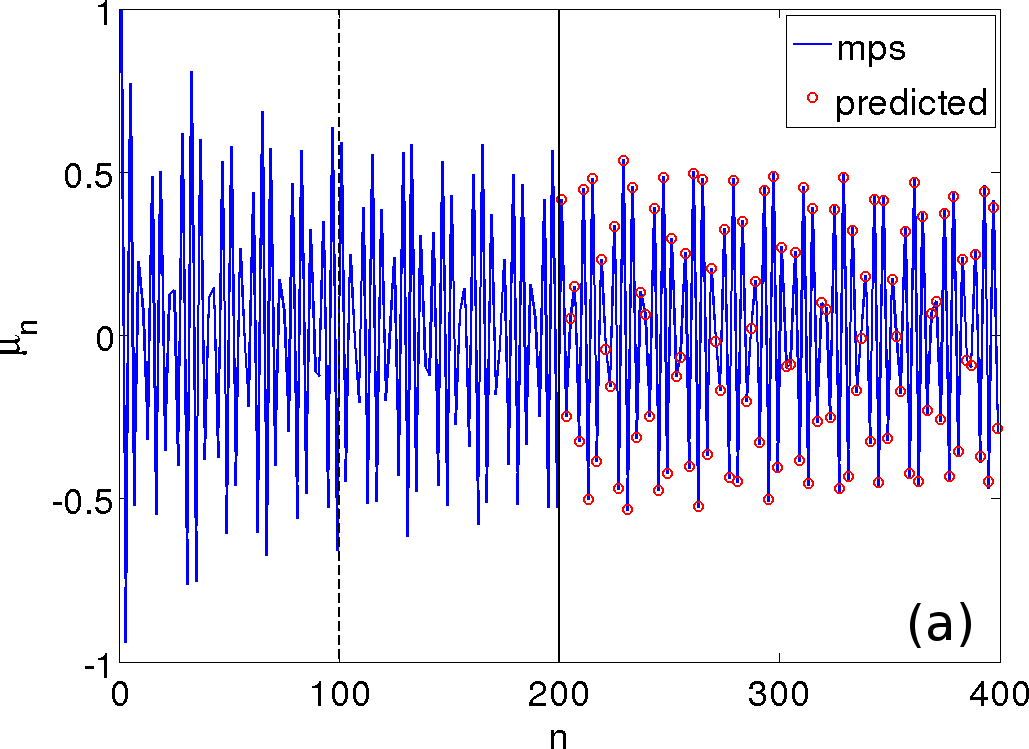}
  \end{minipage}

  \begin{minipage}{1\columnwidth}
    \includegraphics[width=0.98\textwidth]{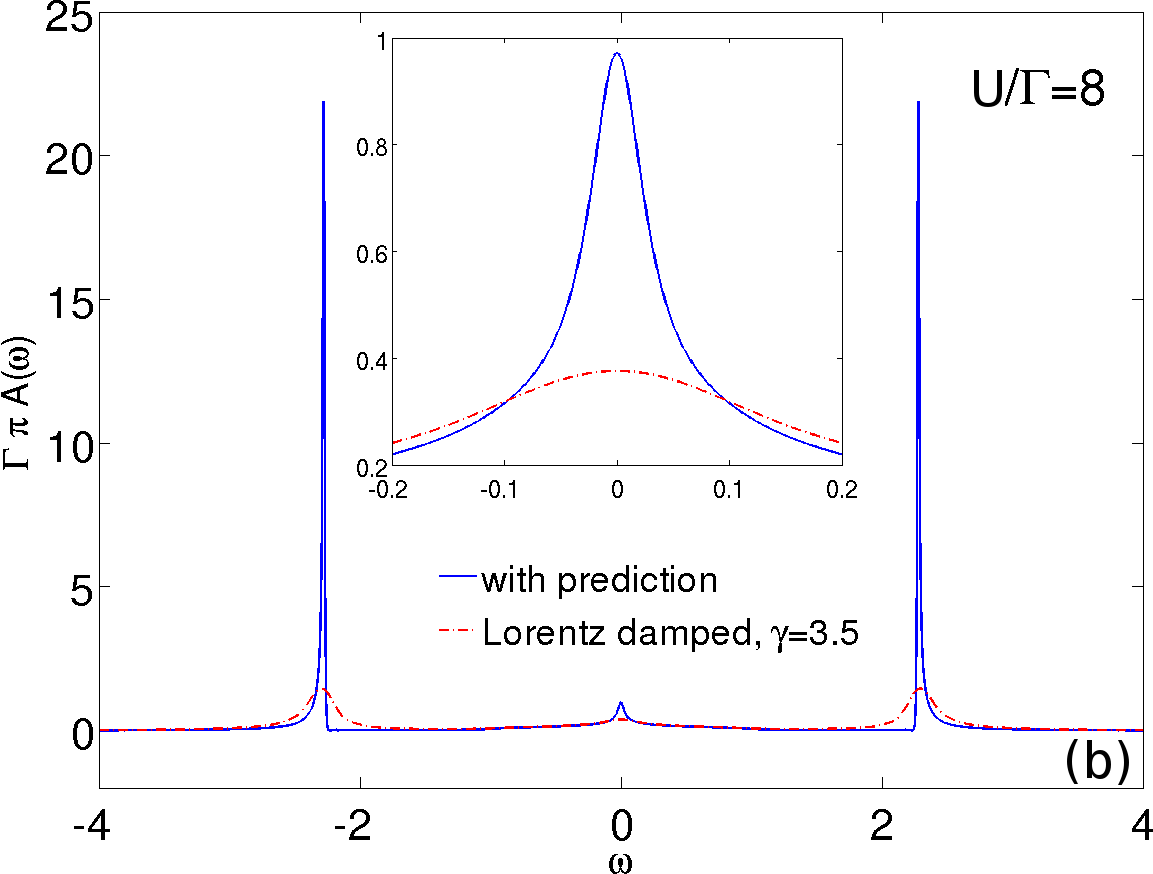}
  \end{minipage}
  \caption{(Color online) 
    (a) Linear prediction using MPS computed Chebyshev moments for the SIAM at $U/\Gamma=8$, $\Gamma=0.5$. 
        Moments to the left of the black dashed line were taken as input data; 
        moments between the dashed and the solid line were used as training-data for the linear prediction algorithm. 
        Only the even moments are plotted. 
    (b) The spectral function (blue line) corresponding to 16400 linearly predicted moments generated from all 400 MPS-computed moments shown in (a).
    This is compared to a Lorentz dampened spectrum using $\gamma=3.5$ and the 400 MPS moments (red dashed line). 
    Inset: magnified region at small frequencies. 
    (Other parameters as in Fig.~\ref{fig:moments_decay}).
\label{fig:comp_damp_vs_pred_u4}}
\end{figure}

%=============================================
While the training moments obtained from the MPS-calculation of the non-interacting RLM in section \ref{sec:rlm} were almost exact, the rapid growth of the truncation errors in the interacting case makes the 
accurate calculation of high-order moments more difficult
and the linear prediction even more important. 
One also needs to consider the effect of truncation errors 
on the moments within the training window of the linear prediction. 
That is, the information gained by adding an additional training moment is offset eventually by its
 numerical error which is passed to the linear prediction. 
When the truncation errors are  small, a large training window can be employed  with excellent result.
In Fig.~\ref{fig:comp_damp_vs_pred_u4}(a), we compare 
MPS computed Chebyshev moments (blue line) with the ones obtained by linear prediction (red circles), where we used the first 200 moments (black solid line) to train prediction.
Fig.~\ref{fig:comp_damp_vs_pred_u4}(b) then shows the spectral function obtained with linear prediction trained on all 400 moments.
For comparison, we show the spectrum obtained by using Lorentz damping 
Eq.~(\ref{eqn:lorentzdamp}) on the original 400 MPS-computed moments,
with damping parameter $\gamma=3.5$ just high enough to remove oscillations. 
The figure clearly demonstrates the increase in spectral resolution achieved by linear prediction.
%--------------------------------------------
%--------------------------------------------

%=============================================
\subsubsection{Comparison with Correction Vector Method (DDMRG)}
\begin{figure} % Fig. 5
\includegraphics[width=1\columnwidth]{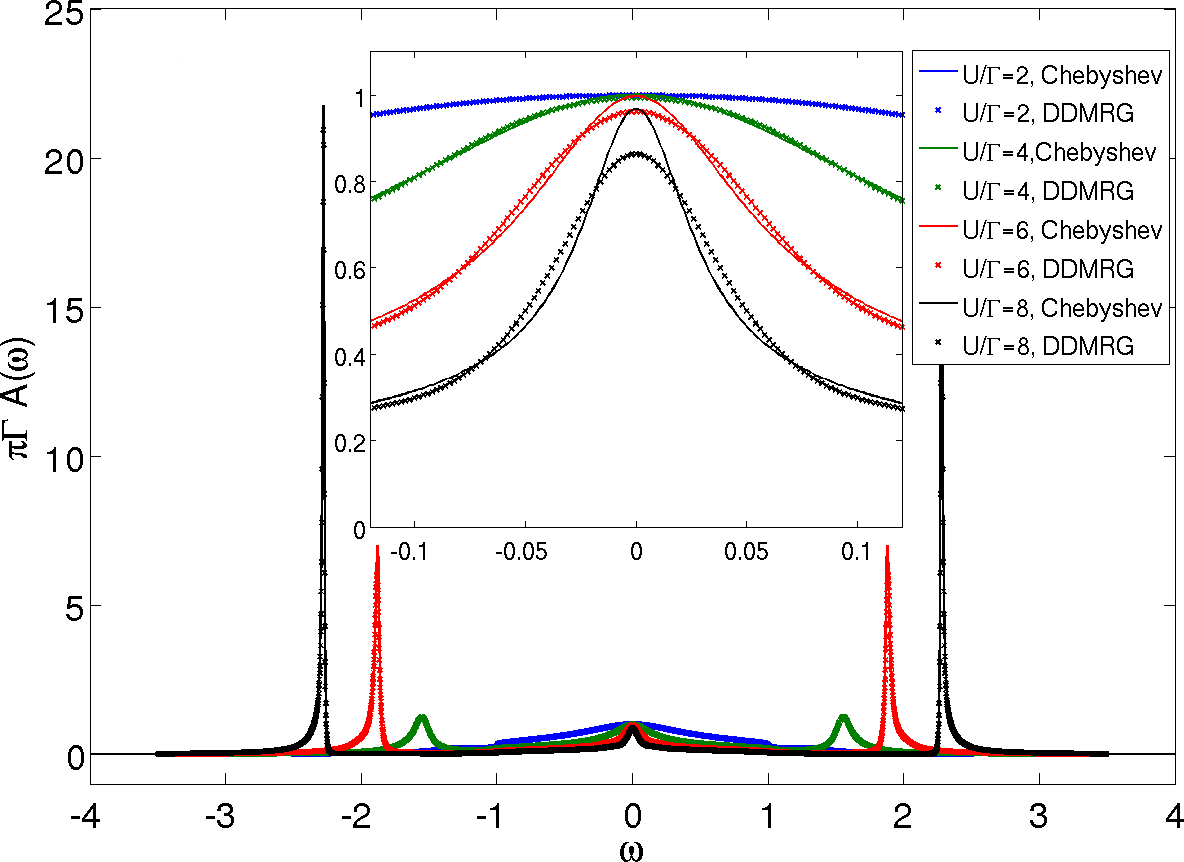}
  \caption{(Color online) 
Impurity spectral function of the SIAM for different values of the interaction $U$, $\Gamma=0.5$.
Solid lines:  spectral function with 400 MPS-computed moments and  16400 further moments from the linear 
    prediction. Symbols:  DDMRG results  \cite{raas_high-energy_2004}  for comparison.
Inset: magnified region at small frequencies. 
The vertical order of lines in the inset is the same as in the key.
(Other parameters as in Fig.~\ref{fig:moments_decay}).
}\label{fig:cheb_vs_cvm}
\end{figure}

%=============================================
The correction vector  (CV) method \cite{kuhner_dynamical_1999} and its variational formulation, the DDMRG \cite{jeckelmann_dynamical_2002}, 
are considered the methods of choice for high precision  calculations of 
dynamical spectral functions of 1-d quantum systems. Their results are assumed to be quasi-exact in many cases. 
Drawbacks of CV (DDMRG) are 
the need for a separate expensive calculation to be done at each frequency $\omega$
and an ill-condition matrix inversion which has to be regularized by a finite (large) broadening of the spectral function,
after which sharp spectral features need to be extracted by a deconvolution procedure.
As a proof of principle, we benchmark our method against results of the CV (DDMRG) \cite{raas_high-energy_2004} in Fig.~\ref{fig:cheb_vs_cvm},
for $U/\Gamma\in \{2,4,6,8\}$. 
Results at $U/\Gamma=8$ are the same as in Fig.~\ref{fig:comp_damp_vs_pred_u4}(b).

We observe the development of  sharp side peaks (Hubbard satellites) 
upon increasing $U/\Gamma$ . The inset of Fig.~\ref{fig:cheb_vs_cvm} shows a zoom onto the zero-frequency region, where with increasing $U$ 
a narrowing of the zero-frequency peak at $U/\Gamma=2$ into a sharp (Kondo) resonance is observed. The agreement with the CV (DDMRG) 
data \cite{raas_high-energy_2004} for $U/\Gamma=2,4$ is excellent. For larger $U/\Gamma=6,8$, we observe deviations in the heights of 
(i) the Hubbard peaks and of (ii) the Kondo resonance. For the latter, the pinning criterion $\Gamma \pi A(0)=1$ \cite{hewson} 
is satisfied to a higher accuracy using Chebyshev expansion with linear prediction than using a maximum entropy  deconvolution of the CV raw data. 
Since the Hubbard satellites are so sharp in this parameter regime, their precise height converges rather slowly with the number of training moments.
%-------------------------------------------
%-------------------------------------------

%=============================================
\subsubsection{Expansion of $\mathbbm{1}-e^{\textrm{-}\tau H}$} \label{sec:1-exp-tauH}
\begin{figure} % Fig. 6
  \includegraphics[width=1\columnwidth]{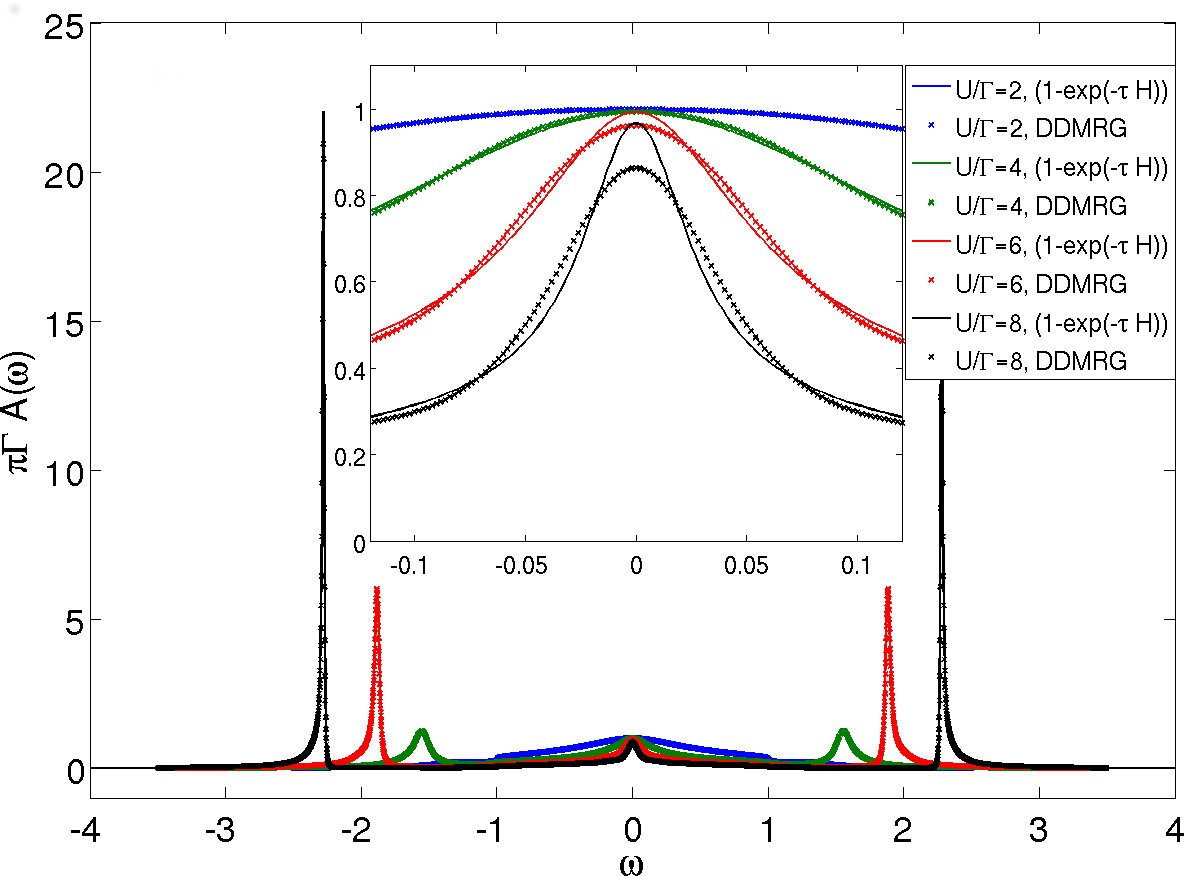}
 \caption{(Color online)
Same as Fig.~\ref{fig:cheb_vs_cvm}, but
using a Chebyshev expansion of $\mathbbm{1}-\exp(\textrm{-}\tau (H-E_0))$ instead of $H$,
with a first order Suzuki-Trotter decoupling ($\tau=0.01,\chi=300$).
We used 1000  ($U/\Gamma=2,4$), 1200 ($U/\Gamma=6$) and 1500 ($U/\Gamma=8$) moments to train the linear prediction ($\delta=10^{-6}$), 
and predicted 20000 ($U/\Gamma=2$), 80000 ($U/\Gamma=4,6$) and 120000 ($U/\Gamma=8$) further moments 
(large number because of small $\tau$). 
}%
\label{fig:expcheb_vs_cvm}
\end{figure}

%=============================================
From the previous discussion, the drawback of the energy truncation scheme \cite{holzner_chebyshev_2011} is the introduction of a systematic error which depends quite strongly
on the choice of auxiliary parameters $a,D_{max}$ and $E_{sweeps}$. In this section, we present first results for the alternate scheme introduced in section \ref{sec:exp_tau}
which employs the expansion of $\mathbbm{1}-e^{-\tau (H-E_0)}$. 
In Fig.~\ref{fig:expcheb_vs_cvm} we compare results 
for $\tau=0.01$ and a first order Trotter expansion of $\exp(\textrm{-}\tau (H-E_0))$ against the same DDMRG data as in Fig.~\ref{fig:cheb_vs_cvm}. 
The results are virtually indistinguishable from those of Fig.~\ref{fig:cheb_vs_cvm}
(except for a very slight difference in the height of the Hubbard peaks),
thus validating our new approach.
When a second order Trotter decomposition is employed, $\tau$ can be increased substantially,
and the required numerical effort should become comparable to that of the energy truncation scheme.

%--------------------------------------------
%--------------------------------------------

%-----------------------------------------
\begin{figure} % Fig. 7 (DMFT)
  \begin{minipage}{1\columnwidth}
    \includegraphics[width=1\textwidth]{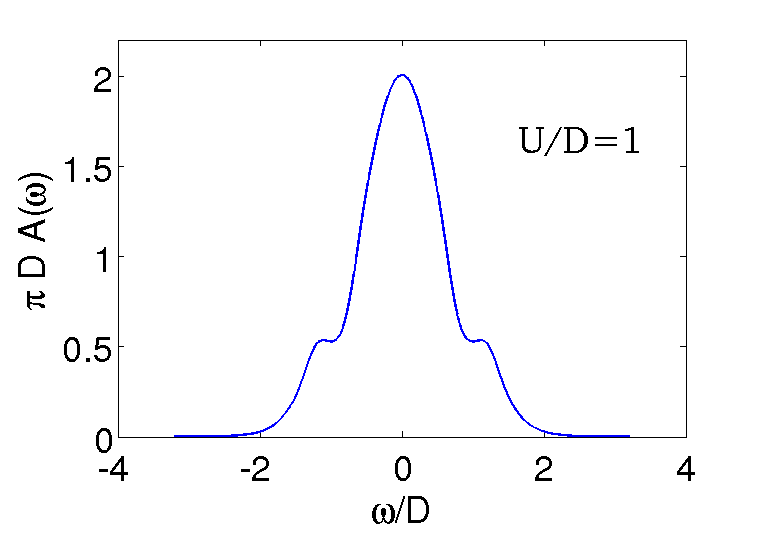}
  \end{minipage}

  \begin{minipage}{1\columnwidth}
    \includegraphics[width=1\textwidth]{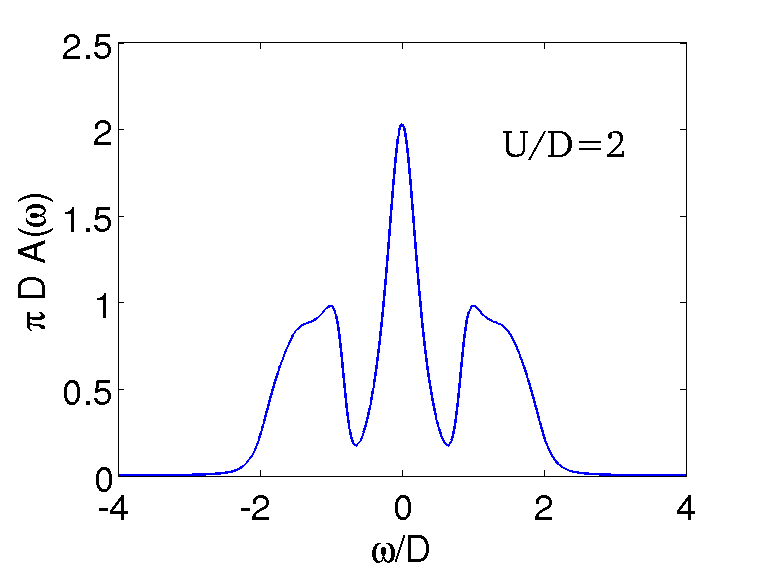}
  \end{minipage}

  \begin{minipage}{1\columnwidth}
\includegraphics[width=1\textwidth]{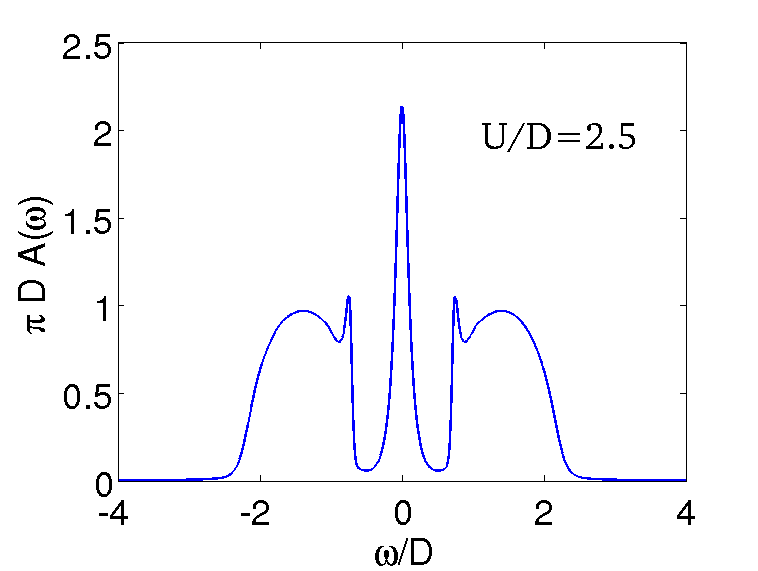}
  \end{minipage}

  \caption{(Color online)
    Local spectral function of the Hubbard model on the Bethe lattice for different interactions $U$.
    With increasing $U$, the formation of Hubbard satellites can be observed. 
    Close to the transition, additional peaks develop at the inner edges of the Hubbard bands.
   (Parameters: upper  panel: $D=1.00$, $a=18$, $D_{max}=5$, $E_{sweeps}=1$, $\chi=250$, $\delta=10^{-6}$, 400 moments calculated; % 200 used for training;
                middle panel: $D=0.25$, $a=12$, $D_{max}=10$, $E_{sweeps}=2$, $\chi=200$, $\delta=10^{-6}$, 600 moments calculated; % 300 used for training;
                lower  panel: $D=0.25$, $a=6$,  $D_{max}=5$,  $E_{sweeps}=1$, $\chi=280$, $\delta=10^{-6}$, 700 moments calculated; % 350 used for training.
}\label{fig:dmft}
\end{figure}
%-----------------------------------------

%=============================================
\subsection{Dynamical Mean-Field Theory}
%=============================================
The DMFT maps the Hubbard model on the Bethe lattice onto an iterative solution of  the SIAM, 
with a hybridization function determined by the impurity Greens function obtained from the previous iteration, 
as shown in Eq. (\ref{eq:dmft_self_con_n}). 
The DMFT scheme provides the exact solution to this model with an infinite number of neighbors once self-consistency has been reached. Note that for obtaining  an accurate DMFT spectrum, additional care is required:
the length of the bath chain $N$ needs to be large enough to avoid finite size artifacts
and to resolve sharp features that are of physical origin. 

In Fig.~\ref{fig:dmft}, we show initial results.
They were obtained with a linear discretization of the bath-DOS with $N=120$ sites
 We see a narrowing of the quasiparticle peak at $\omega=0$ with increasing interaction, and the formation
of Hubbard satellites at $\omega\approx \pm U/2$ \cite{georges_dynamical_1996,karski_single-particle_2008,garcia_dynamical_2004,zitko_energy_resolution_2009,osolin_pade_2013,nishimoto_dynamical_2004}. 
For  $U/D=2.5$, an additional peak in the Hubbard band can be clearly identified. 
In previous NRG \cite{Bulla1999}
 studies, this peak could not be resolved.
Our results are compatible with  studies using improved resolution NRG \cite{zitko_energy_resolution_2009},
and DDMRG \cite{karski_single-particle_2008}
in which such a peak has been seen,  albeit with conflicting results regarding its sharpness.
\\

%=============================================
\section{Conclusions}
%=============================================

We proposed two extensions of a recently developed MPS-based method for expanding spectral functions in Chebyshev 
polynomials \cite{holzner_chebyshev_2011}. We used the linear prediction algorithm to extrapolate moments up to 
high orders, which  significantly improved the achievable resolution at practically no computational cost.
This is especially interesting in systems where strong growth of site-entanglement (bipartite entanglement entropy) 
prevents one from iterating the recursion to high orders, due to increasing truncation effects. 
We benchmarked the method with the exactly solvable resonating level model, where we obtained highly accurate results. 
We also investigated the single impurity Anderson model and obtained results which compare very well with spectra obtained
from the correction vector method (CV, DDMRG) \cite{raas_high-energy_2004,raas_spectral_2004}, at significantly 
reduced computational cost \cite{holzner_chebyshev_2011}.
We further applied the method
as a high resolution impurity solver within dynamical mean field theory \cite{georges_dynamical_1996}. 
Particular advantages are that 
(i) the method works at zero temperature and on the real frequency axis, 
(ii) it works for an {\it arbitrary} discretization grid of the bath density of states (different from  NRG), which allows for good energy resolution at all frequencies, and 
(iii) it is applicable to any 1-d model with short range interaction. 
Results confirmed the existence of pronounced peaks at the inner edges of the 
Hubbard bands in the metallic phase of the Hubbard model. To overcome the shortcomings of energy truncation of 
the Chebyshev MPS-method (\cite{holzner_chebyshev_2011}, we proposed a modified rescaling scheme which employs a 
Chebyshev expansion of $\mathbbm{1}-\exp(\textrm{-}\tau (H-E_0))$, for which the energy truncation 
step \cite{holzner_chebyshev_2011} can be {\it completely omitted}, at a comparable spectral resolution.
The implementation of the scheme is very similar to standard time evolution algorithms
\cite{vidal_efficient_2004,vidal_efficient_classical_2004,daley_time-dependent_2004, white_real-time_2004}.

Both methods are promising candidates for high resolution, low $T$ impurity solvers for DMFT.
Whereas in NRG more than two orbitals become computationally too demanding,
extensions to multi-orbital systems and finite temperatures are within reach of our approach.

\begin{acknowledgments}
We acknowledge financial support by the Austrian Science Fund through SFB ViCoM F41 P03 and P04. 
Calculations have been done in part on the Vienna Scientific Cluster.
We would like to thank S. Andergassen for interesting discussions, 
and C. Raas for providing his DDMRG data shown in Figs.~\ref{fig:cheb_vs_cvm} and \ref{fig:expcheb_vs_cvm}.
\end{acknowledgments}

%-----------------------------
\begin{figure}[tb] % Fig. 8
\includegraphics[width=1\columnwidth]{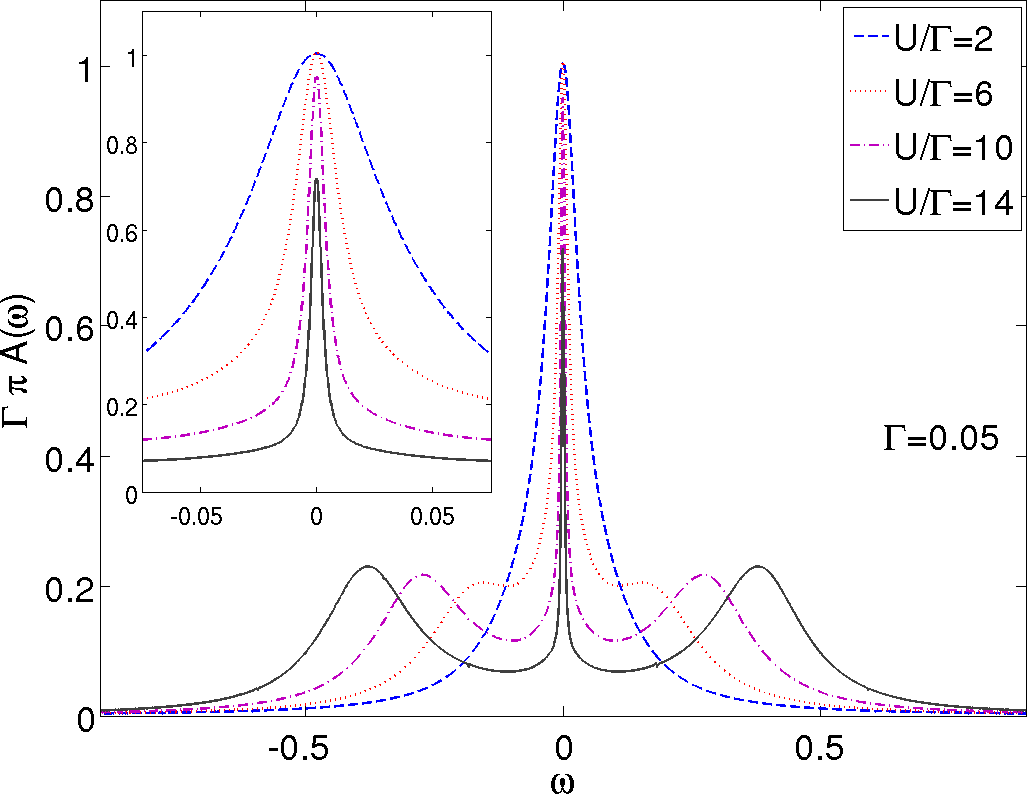}
  \caption{(Color online)
Spectral function of the SIAM for $U/\Gamma = 2,6,10,14$ in the wide band regime ($\Gamma=0.05, D=1$).
In the central region, one observes a successive narrowing of the zero-frequency peak 
which results in the Kondo-resonance. 
 The outer Hubbard satellites with peak position at $\approx U/2$ are also clearly visible. 
The inset shows a zoom onto the zero frequency region. 
(MPS parameters: $\chi=180, a=5, D_{max}=30$, $E_{sweeps}=5$).
}\label{fig:siam_cheby}
\end{figure}
%-----------------------------

%--------------------------------------------
\begin{figure*}[tb] % Fig. 9
%--------------------------------------------
  \begin{minipage}{1\textwidth}
    \includegraphics[width=0.47\textwidth]{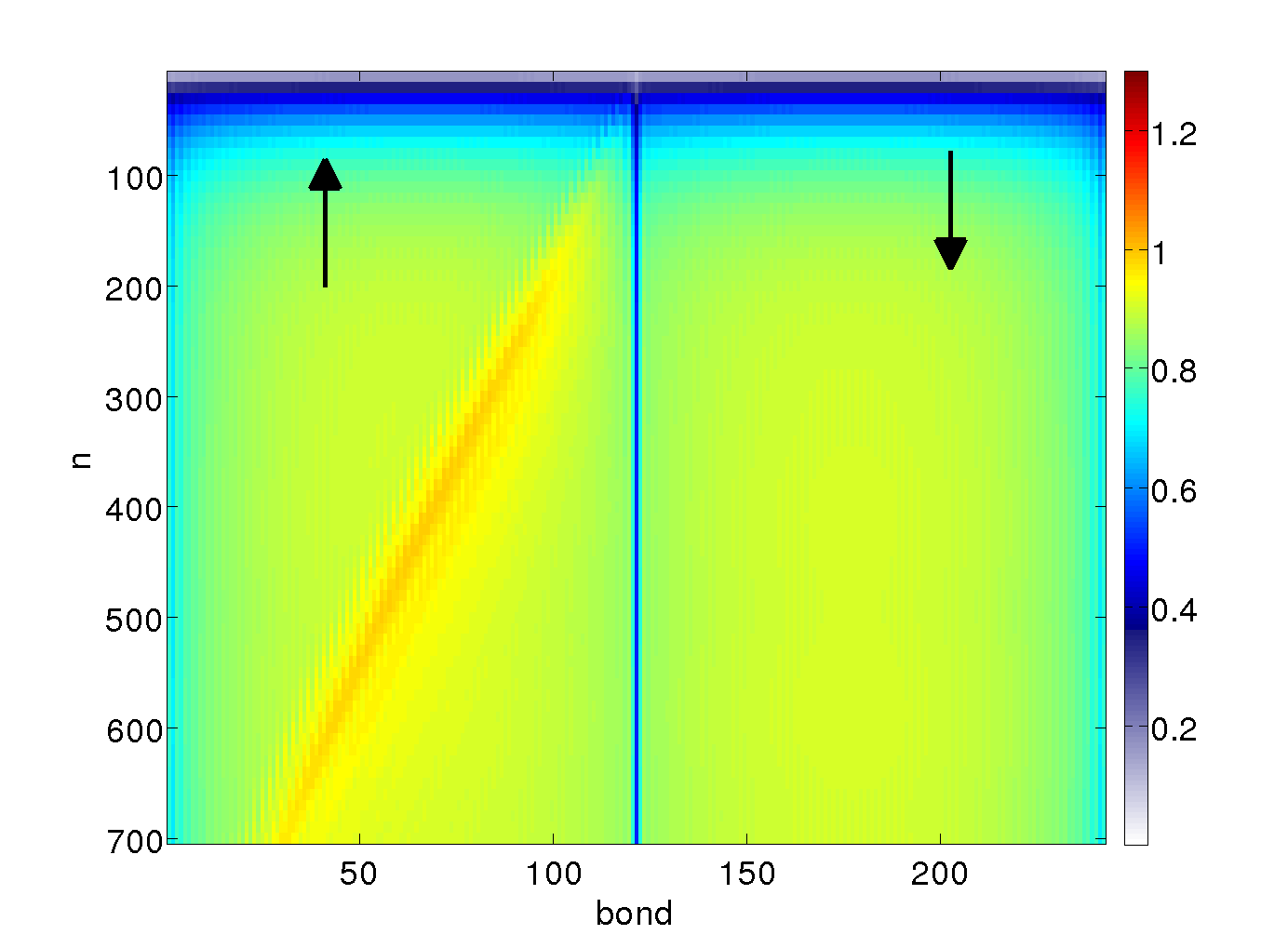}
    \includegraphics[width=0.51\textwidth]{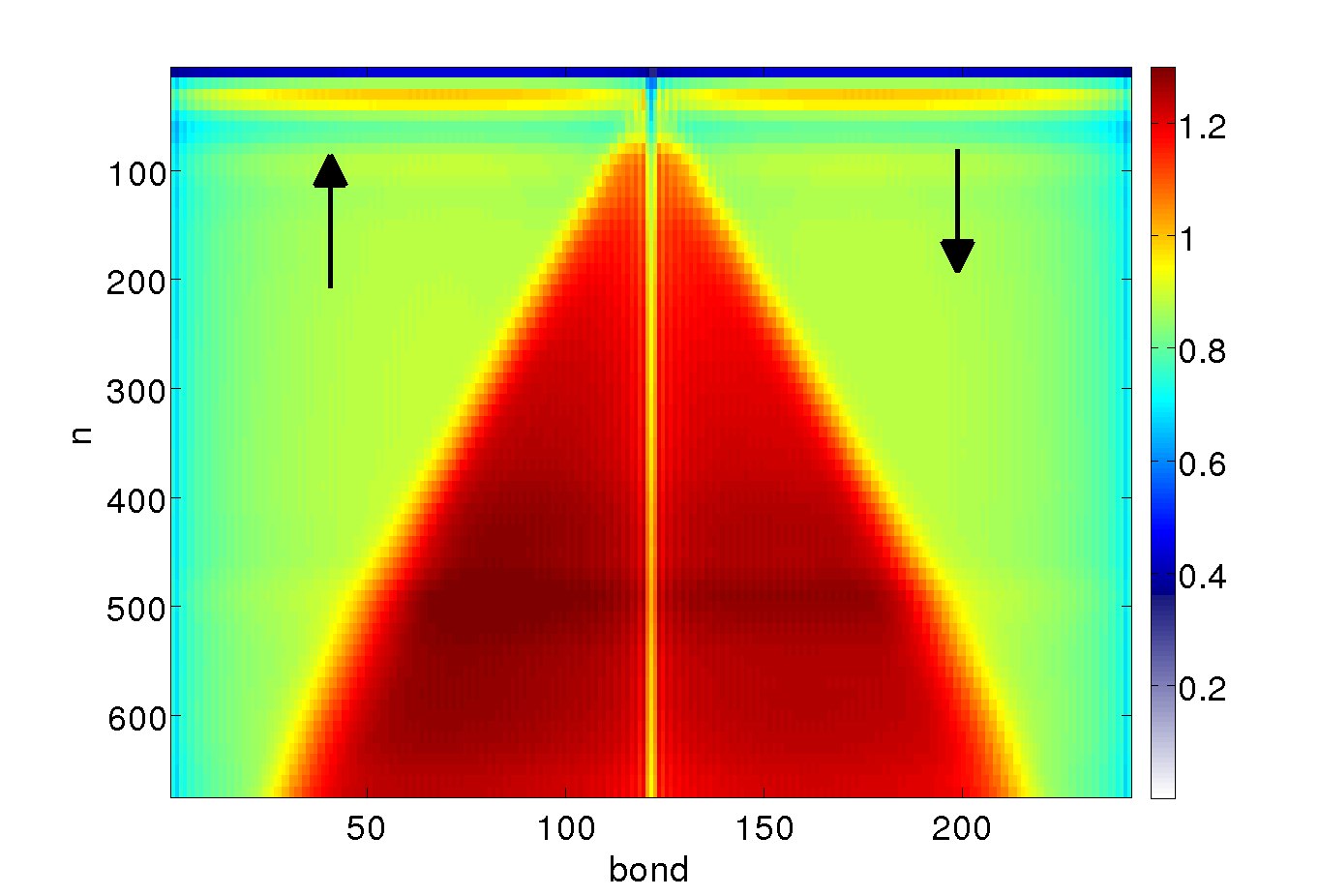}
  \end{minipage}
\caption{(Color online)
Bipartite entanglement
at different bonds (x-axis)  of the state $\ket{t_n}$ ($n$ on y-axis) obtained during calculation of the Chebyshev moments for the {\em up}-spin,
 particle branch of the impurity spectral function
of the SIAM (i.e.\  $\ket{t_0}=c_{\uparrow}^{\dagger}\ket{\Psi_0}$). 
The impurity is located at bond 120, 
with up-spins to the left and down-spins to the right.
Left panel:  RLM ($U=0$). Right panel: SIAM for $U=0.5$. (In both plots, $\Gamma=0.05$, $a=6$, $N=120$, $\chi=300$, $D_{max}=5$, $E_{sweeps}=1$.)
%(semicircular bath DOS with $D=1$ and mixed log-linear discretization
}
\label{fig:entropy}
\end{figure*} 
%---------------------------------

%=============================================
\section*{Appendix A: Wide rectangular bath DOS \label{app:log}}
%=============================================

Here we examine the case of a wide band ($U \ll D$) and focus on
the low energy scale associated with the Kondo resonance. A logarithmic discretization mesh, $x_n=\Lambda^{-n}$ ($\Lambda$ = 1.05), with a chain size of N = 100 is used to resolve the sharp resonance. The results are obtained for a flat conduction band 
\begin{equation}
  \rho(\omega) = \begin{cases} 1/(2D), & \omega \in [-D,D] \\ 0 & \mbox{else} \end{cases},\label{eq:flat_dos}
\end{equation}
with bandwidth $2D=2$, $\epsilon_f=-U/2$ (particle-hole symmetric point), and a hybridization strength $\Gamma=0.05$. 

Fig.~\ref{fig:siam_cheby} shows impurity spectral functions for $U/\Gamma=2,6,10,14$. With increasing $U$, one observes a narrowing of the central conduction peak, accompanied by the formation of Hubbard 
satellites at $\omega\approx U/2$. Note the different parameter regime as compared to Fig.~\ref{fig:cheb_vs_cvm}, where the Hubbard satellites lie well outside the bandwidth of the bath. Now, in the 
wide bandwidth regime $U\ll D$, the Hubbard satellites are much broader.  
The inset shows a zoom onto the low-frequency region. Besides the 
narrowing of the Abrikosov-Suhl resonance, we observe that with increasing $U$, the pinning criterion is no more obeyed. This is not unexpected, since the lifetime of the quasiparticle scales inversely
with the resonance width, leading to an exponential increase in the Chebychev expansion order needed to resolve this resonance. Using linear prediction increases the achievable resolution, but results of course also 
depend on the size of the training set as well as the accuracy of the data. If the set is too small, so that signatures of the resonance are not strong enough to be picked up properly by prediction, 
it is not fully resolved by the method.
Indeed, the height at $\omega=0$ is sensitive to parameters like the size of the training window and the cutoff $\delta$ for inversion.
In some cases, it can vary by 20-30\%.
The exact form of the resonance also depends on the discretization of the band around $\omega=0$. 
If the discretization is too crude, we observe in general an underestimation of the height of the resonance.

The Hubbard peaks, on the other hand, are not sensitive at all.
Importantly, while the precise height at $\omega=0$ can be sensitive to parameters of the calculation,
we observe that the {\em weight} of the resonance, i.e.\ the integral over the resonance peak, is very stable.
\\%

%=============================================
\section*{Appendix B: Entanglement and truncated weight growth\label{app:ent}}
%=============================================

Time scales for MPS simulations are usually limited by the growth of site-entanglement between the separate parts of the quantum system. One (non-unique) way of quantifying site-entanglement is the bipartite entanglement
entropy $S_{vN}=-{\rm tr}_B\left(\rho_B{\rm log}\rho_B\right)$ \cite{eisert_area_2008}, with $\rho_B={\rm tr}_A\rho_{AB}$. $\rho_{AB}$ is the full density matrix of a bipartite quantum system $A,B$, and ${\rm tr}_A$ denotes the partial trace 
over all degrees of freedom in part $A$ of the system. Using MPS with a maximum bond-dimension amounts to essentially introducing an upper bound $\sim \log \chi_{max}$ to $S_{vN}$. 
The error of this approximation
can be quantified by the truncated weight 
\begin{equation}
  \epsilon_{tw}=1-\sum_{i=1}^{\chi}\lambda_i^2 \; ,
\end{equation}
where $\lambda_i$ denote the Schmidt-coefficients \cite{nielsen-chuang} belonging to the bipartition $A:B$ 
(i.e.\ $\lambda_i^2$ are the simultaneous eigenvalues of $\rho_{A}$ and $\rho_B$), and $\chi$ is the matrix dimension of the MPS-matrices.
In our simulations we observed a strong increase of truncated weight which limits the number of computable moments. In Fig.~\ref{fig:entropy}, we plot the entanglement entropy
for the states $\ket{t_n}$ obtained during the Chebyshev expansion of the positive up-spin part of the spectral-function of the SIAM (i.e. $\ket{t_0}=c_{\uparrow}^{\dagger}\ket{\Psi_0}$).
The left panel shows results for $U=0,\Gamma=0.05$ and a semicircular bath DOS with $D=1$, discretized into  N$=120$ sites (impurity included).
Due to the unfolding, the left side of the plot represents the up spins and the right side the down spins. The added up-spin particle thus travels along the chain and locally increases 
entanglement around its position. However, after the particle has passed a certain bond, entropy again decreases. Importantly, the signal travels {\it only in the up-spin branch}, due to the missing entanglement between up- and down spins in the initial ground state $\ket{\Psi_0}$. 
The truncated weight for this simulation never exceedes 1e-6.

The right panel in Fig.~\ref{fig:entropy} shows the same plot for finite $U=0.5$. Again, we observe a propagating signal, but this time it spreads in both directions, e.g. in the up {\it and} down 
spin channel. Furthermore, after passage of the signal at a certain bond, the entropy increases and remains at this higher value. Both effects are due to the presence of strong correlations in the initial state. 
Oscillations on top of the signal are due to the change in norm of $\ket{t_n}$ during the simulation. 
For $n>300$ the truncated weight already exceeds a value of 1e-3, and only 
results for $n<300$ should be considered as reliable in this simulation.

\end{document}